\documentclass[%
 aip,
 chaos,%
 amsmath,amssymb,
 reprint,%
]{revtex4-1}

\usepackage{graphicx}

\usepackage{color}

\begin{document}

\title{Brownian ratchets: How stronger thermal noise can reduce diffusion}

\author{Jakub Spiechowicz}
\affiliation{Institute of Physics, University of Silesia, 40-007 Katowice, Poland}
\affiliation{Silesian Center for Education and Interdisciplinary Research, University of Silesia, 41-500 Chorz{\'o}w, Poland}

\author{Marcin Kostur}
\affiliation{Institute of Physics, University of Silesia, 40-007 Katowice, Poland}
\affiliation{Silesian Center for Education and Interdisciplinary Research, University of Silesia, 41-500 Chorz{\'o}w, Poland}

\author{Jerzy {\L}uczka}
\email{jerzy.luczka@us.edu.pl}
\affiliation{Institute of Physics, University of Silesia, 40-007 Katowice, Poland}
\affiliation{Silesian Center for Education and Interdisciplinary Research, University of Silesia, 41-500 Chorz{\'o}w, Poland}


\begin{abstract}
We study  diffusion properties of an inertial Brownian motor moving on a ratchet substrate, i.e. a periodic structure with broken reflection symmetry. The motor is driven by  an unbiased time-periodic symmetric force which takes the system out of thermal equilibrium. For selected parameter sets, the system is in a non-chaotic regime in which we can identify a non-monotonic dependence of the diffusion coefficient on temperature: for low temperature, it initially increases as temperature grows, passes through
its local maximum, next starts to diminish reaching its
local minimum and finally it monotonically increases in accordance with the Einstein linear relation. Particularly interesting is the temperature interval in which diffusion is suppressed by thermal noise and we explain this effect in terms of transition rates of a three-state stochastic model.
\end{abstract}


\maketitle

\begin{quotation}
At the meso- and nanoscale transport of Brownian particles under   nonequilibrium conditions can exhibit  features which are very different from those observed in the macroscopic world. At such  scales, influence of thermal noise may be 
constructive to the dynamics rather than playing the usual
destructive role.  
 Despite the fact that many theoretical frameworks have been developed our current understanding of nonequilibrium physics fundamentals  still remains incomplete, undoubtedly far beyond what we know for equilibrium systems.  Recent advances in the field are comprehension of stochastic resonance, noise-assisted transport in ratchet systems, negative mobility, anomalous diffusion and fluctuation theorems. Here we present another fascinating manifest of such systems, namely the phenomenon of non-monotonic temperature dependence of diffusion which is strictly ruled out in equilibrium by the famous Einstein relation. We expound the mechanism standing behind this peculiar behaviour which can be realized both in classical and quantum setups, in solid state physics as well as in soft and active matter systems, in physical, chemical and biological systems.
\end{quotation}

\section{Introduction}
Diffusion is a universal phenomenon observed in diverse systems. It plays a crucial role not only in physical, chemical or biological setups but also its concept is used in socio-economical contexts in such processes like diffusion of ideas or innovations \cite{rogers2003}. Over one hundred years ago Einstein \cite{einstein1905} and Smoluchowski \cite{smoluchowski1906} formulated the theory of Brownian motion which provides a link between the microscopic dynamics and the macroscopically observable diffusion. According to it, in thermal equilibrium the spreading of a cloud of independent  Brownian particles is more effective at higher temperature. But can diffusion even decrease with temperature? After all, Nature is prodigal in presenting sophisticated mechanisms that regulate phenomena which takes place in all scales of time and space. These control strategies reach the ultimate level of efficiency and refinement in biological systems which are responsible for the emergence of the sustainable phenomenon of Life \cite{chowdhury2013,spiechowicz2013jstatmech, spiechowicz2014pre,spiechowicz2016jstatmech}. A prominent example may be an intracellular transport mediated by molecular motors \cite{bressloff2013, hanggi2009}. It is an archetypal system, in which, even in the absence of externally applied bias, directed motion emerges by harvesting thermal fluctuations via the mechanism of breaking of spatio-temporal symmetries of the setup \cite{hoffmann2016}.

In view of the example considered in this paper we address the question whether it is possible to observe diffusion decreasing with temperature for the system far from thermal equilibrium. This peculiar behaviour should be clearly contrasted with the already renowned case of anomalous diffusion \cite{metzler2014, zaburdaev2015} or the phenomenon of giant diffusion \cite{lindner2001, heinsalu2004, lindner2016, dan2002, garcia2014}. To unravel the posed problem we study nonequilibrium dynamics of a Brownian motor. The importance of the latter in science has become evident in the last two decades due to its widespread applications in both biological and non-biological, artificial systems  \cite{vlassiouk2007, serreli2007, mahmud2009, costache2010, drexler2013, spiechowicz2014prb, spiechowicz2015njp, spiechowicz2015chaos, roche2015, grossert2016}. We demonstrate that the diffusion coefficient $D$ may be a non-monotonic function of temperature $\theta$. Initially $D$ increases as $\theta$ grows, passes through its local maximum and next starts to decrease reaching its local minimum to get larger with it later again. This counter-intuitive behaviour does not look like an exception and has been found in a variety of setups including cytoplasmic protein diffusion \cite{guo2014}, zeolite-guest systems \cite{schuring2002}, polymer nanocomposites \cite{tung2016}, $^3He-^4He$ mixtures at low temperature \cite{ganshin1999}, nonuniform twisted vortex states in rotating superfluids \cite{eltsov2006}, quasiparticles coupled to vibrations of a one-dimensional non-linear chain of atoms \cite{iubini2015} and finally extended disordered systems in contact with the phonon bath over complete range of a dissipation strength \cite{lee2015}. However, in this paper with our relatively simple and clear model of a Brownian motor we aim to explain the mechanism standing behind the mentioned peculiar diffusive behaviour. Our finding is in some sense universal and can be realized both in classical and quantum setups; in solid state physics as well as in soft matter; in physical, chemical and biological systems.
 
The layout of the present paper is organized as follows. In Section II we describe in detail a model of the periodically driven Brownian motor moving on the asymmetric substrate and introduce all quantities of interest. In Section III, its deterministic counterpart is briefly analysed. For the considered parameter set,  the system is non-chaotic and possesses  three paramount attractors. Next, in Section IV, we show that for this  regime, a  non-monotonic temperature dependence of the diffusion coefficient is observed. In Section V we consider the distribution of the period averaged velocity in the regime of long time. Its three maxima are located around the positions of three deterministic paramount attractors. These peaks can constitute a three-state stochastic model with jumps induced by thermal equilibrium fluctuations. In Section VI  we consider  transition probabilities between three states as a function of temperature  of the system. It allows to explain the observed peculiar diffusive behaviour. To better visualize the proposed mechanism, in Section VII we study  spread of  trajectories of the Brownian motor.  Finally, in  Section VIII, we conclude the paper with a discussion and a summary of the main findings.
\begin{figure}[t]
	\centering
	\includegraphics[width=1.0\linewidth]{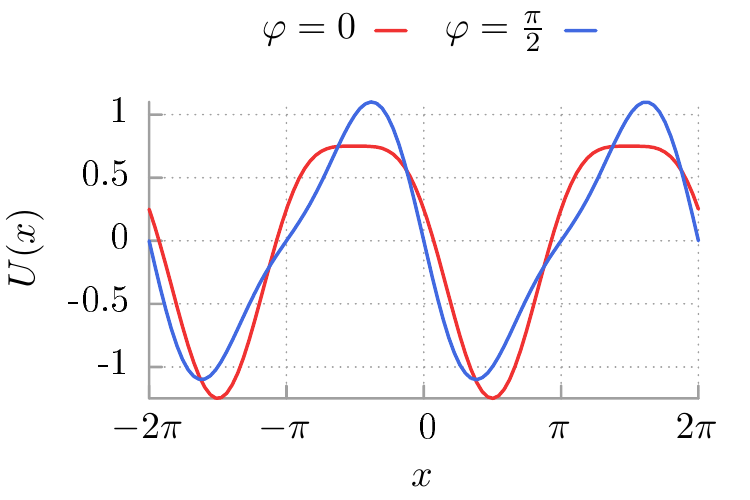}
	\caption{The potential given by Eq. (\ref{potential}) depicted in the symmetric case $\varphi = 0$ in comparison with a ratchet form for the asymmetry parameter $\varphi = \pi/2$.}
	\label{fig1}
\end{figure}
\section{Brownian motor in ratchet potential}
We consider an archetypal model of a Brownian motor consisting of a classical inertial particle of mass $m$, which moves in a spatially periodic ratchet-type potential $U(x)$, see \mbox{Fig. \ref{fig1}}. It is driven by an unbiased time periodic force $a\cos{(\omega t)}$ of amplitude $a$ and angular frequency $\omega$ and subjected to thermal noise of intensity $Q$. Dynamics of the particle is determined by the Langevin equation, which in the dimensionless form reads
\begin{equation}
	\label{model}
	m\ddot{x} + \dot{x} = -U'(x) + a\cos{(\omega t)} + \sqrt{2Q}\,\xi(t).
\end{equation}
The dot and the prime denote differentiation with respect to time $t$ and the particle coordinate $x \equiv x(t)$, respectively. The dimensionless noise intensity $Q = k_B \theta/\Delta U$ is given by the ratio of thermal energy
$k_B \theta$ and a half of the activation energy the particle needs to overcome the original potential barrier $2\Delta U$. The parameter $k_B$ is the Boltzmann constant.
We refer the reader to Ref. 23 for a complete overview of the scaling of the above equation. Thermal fluctuations are modelled by $\delta$-correlated Gaussian white noise of zero mean and unit intensity, i.e.
\begin{equation}
	\label{noise}
	\langle \xi(t) \rangle = 0, \quad \langle \xi(t)\xi(s) \rangle = \delta(t-s).
\end{equation}
The ratchet potential $U(x)$ is assumed to be in the following form \cite{spiechowicz2015pre}
\begin{equation}
	\label{potential}
	U(x) = -\sin{x} - \frac{1}{4} \sin{\left( 2x + \varphi - \frac{\pi}{2}  \right)},
\end{equation} 
where the relative phase $\varphi$ between  two harmonics serves as a control parameter of the reflection-symmetry of this potential. It is reflection-symmetric when there exists a shift $x_0$ such that $U(x_0 + x) = U(x_0 - x)$ for all $x$. If $\varphi \neq 0, \pi$ then the mirror symmetry is broken which we in turn classify as a ratchet potential. The system described by Eq. (\ref{model}) is not a toy model but has a wealth of physical realizations including, among others \cite{hanggi2009}, an asymmetric superconducting quantum interference device (SQUID) which is composed of three capacitively and resistively shunted Josephson junctions \cite{spiechowicz2014prb, spiechowicz2015njp, spiechowicz2015pre}. Two junctions are collocated in series in one half-piece of the arm and the third junction is disposed in the other half of the arm. The SQUID is subjected to a time-periodic current $a\cos{(\omega t)}$ and pierced by an external constant magnetic field proportional to $\varphi$. In consequence, asymmetry of the potential can be controlled by the external magnetic field.

We study a diffusion process of the particle position $x(t)$ and the spread of trajectories. In literature one can find several similar quantifiers which characterize this phenomenon. One of them is the mean square deviation (or variance) of the coordinate $x(t)$ from its average, namely, 
\begin{equation} 
	\label{Delta}
	 \langle \Delta x^2(t) \rangle = \langle [ x(t) - \langle x (t) \rangle ]^2 \rangle = \langle x^2(t) \rangle - \langle x(t) \rangle^2,
\end{equation}
where averaging is over all possible thermal noise realizations as well as over initial conditions for the position $x(0)$ and velocity $\dot x(0)$ of the Brownian motor. In many cases, in the asymptotic long time regime it grows according to a power law \cite{metzler2014,zaburdaev2015}
\begin{equation}
	\langle \Delta x^2(t) \rangle \sim t^{\alpha}.
\end{equation}
The exponent $\alpha$ determines a type of diffusion: normal diffusion is when $\alpha = 1$, subdiffusion is developed for $0 < \alpha < 1$ and superdiffusion occurs for $\alpha > 1$. Only when $\alpha = 1$ the time-independent diffusion coefficient $D$ can be computed as 
\begin{equation}
\label{D}
	D = \lim_{t \to \infty} \frac{\langle [ x(t) - \langle x(t) \rangle ]^2 \rangle}{2t}. 
\end{equation}  
Otherwise the above definition is not constructive because such a quantity is either zero (subdiffusion) or diverges to infinity (superdiffusion). 

The stochastic process $x(t)$ determined by Eq. (\ref{model}) exhibits 
various forms of diffusion anomalies.  In previous papers \cite{spiechowicz2015pre, spiechowicz2016scirep} it was shown that for finite times,   transient anomalous effects  can occur. In particular, the mean square deviation 
$\langle \Delta x^2(t) \rangle$  initially evolves in a superdiffusive way, next subdiffusion is observed and finally it approaches normal diffusion behavior.  It is worth to stress that lifetimes of the superdiffusion and subdiffusion  can be many, many orders
of magnitude longer than the characteristic time scale of the system and turns out to be extraordinarily sensitive to the system parameters like temperature or the potential asymmetry \cite{spiechowicz2016scirep}. In the the asymptotic long time limit $t\to\infty$, diffusion is always {\it normal}. In this paper, we study normal diffusion in the stationary states 
for $t\to\infty$ with the well-defined  diffusion coefficient $D$ determined by the relation (\ref{D})  and analyze the non-monotonic dependence of $D$ on temperature. Moreover, we reveal the mechanism standing behind the emergence of this phenomenon.   
 
All three elements entering the right hand side of Eq. (\ref{model}) are unbiased: the average of the potential force $-U'(x)$ over the spatial period $L = 2\pi$ vanishes as well as that of the time dependent driving $a\cos{(\omega t)}$ over a temporal period $T = 2\pi/\omega$ and also the mean value of the random force $\xi(t)$ vanishes according to Eq. (\ref{noise}). However, due to the presence of the external driving the Brownian particle is taken far away from thermal equilibrium and a time dependent nonequilibrium state is reached in the asymptotic long time regime. Then the mean velocity $\langle \dot{x} \rangle$ takes the form of a Fourier series over all possible harmonics \cite{jung1993}
\begin{equation}
	\lim_{t \to \infty} \langle \dot{x}(t) \rangle = \langle \mathbf{v} \rangle + v_\omega(t) + v_{2\omega}(t) + ...,
\end{equation}
where $\langle \mathbf{v} \rangle$ is the directed (time independent) transport velocity while $v_{n\omega}(t)$ denote time periodic higher harmonic functions of vanishing average over the fundamental period $T = 2\pi/\omega$. In our setup, a necessary condition for the occurrence of directed transport $\langle \mathbf{v} \rangle \neq 0$ is the breaking of the reflection symmetry of the potential $U(x)$ \cite{hanggi2009} which occurs for $\varphi \neq 0, \pi$.  Due to this particular decomposition of the asymptotic long time average velocity it is useful to study also the period averaged velocity $\mathbf{v}(t)$ defined as
\begin{equation}
	\mathbf{v}(t) = \frac{1}{T} \int_t^{t + T} ds\,\dot{x}(s)
\end{equation}
which may be utilized to compute the directed transport velocity in the following  way
\begin{equation}
	\langle \mathbf{v} \rangle = \lim_{t \to \infty} \langle \mathbf{v}(t) \rangle.
\end{equation}

The Fokker-Planck equation corresponding to the  model (\ref{model}) cannot be solved by any known analytical methods. Therefore we performed comprehensive numerical simulations of the driven Langevin dynamics. We did it by using a weak version of the stochastic second-order predictor-corrector algorithm with a time step typically set to $10^{-2} \times T$. Our quantities of interest we averaged over $10^4$ sample trajectories as well as over initial conditions $x(0)$ and $\dot{x}(0)$ equally distributed in the intervals $[0,2\pi]$ and $[-2,2]$, respectively. Numerical calculations have been performed with a CUDA environment implemented on a modern desktop GPU. This procedure did allow for a speedup of a factor of the order $10^3$ times as compared to a common present-day CPU method, for details see Ref. 38.
\begin{figure}[t]
	\centering
	\includegraphics[width=1.0\linewidth]{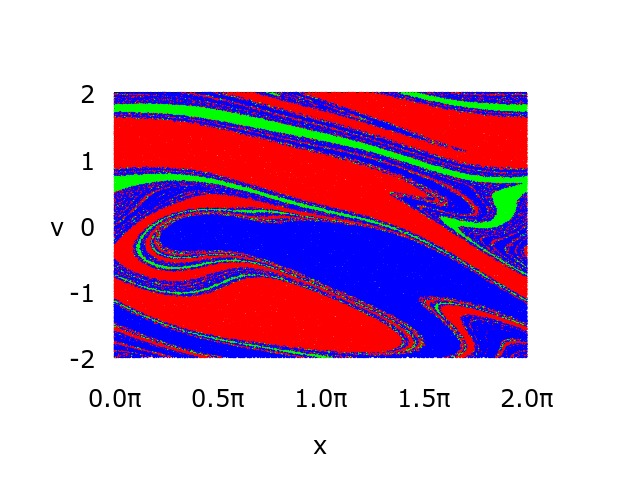}
		\caption{Basins of attraction for the asymptotic long time particle velocity $\mathbf{v}(t)$. The red and blue coloured set consists of all initial conditions $\{x(0), \dot{x}(0)\}$ eventually evolving to the running states with the positive $v_+ = \mathbf{v}(t) \approx 0.4$ and negative $v_- = \mathbf{v}(t) \approx -0.4$ velocity, respectively. The green colour set marks the locked states $v_0 = \mathbf{v}(t) \approx 0$. Parameters are: $m = 6$, $a = 1.899$, $\omega = 0.403$, $\varphi = \pi/2$. For this particular regime the deterministic system (\ref{model}) with $Q = 0$ is non-chaotic.}
		\label{fig2}
\end{figure}
\begin{figure}[t]
	\centering
	\includegraphics[width=1.0\linewidth]{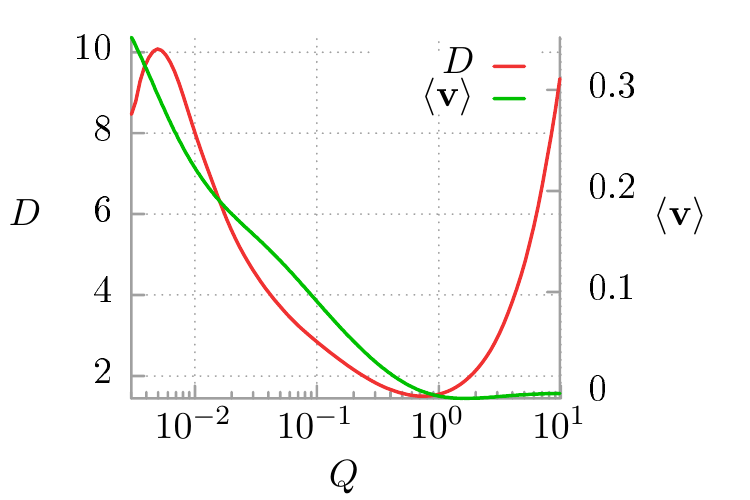}
	\caption{The dependence of the diffusion coefficient $D$ and the directed (time independent) transport velocity $\langle \mathbf{v} \rangle$ on temperature $Q$ of the system. $D$ displays the non-monotonic behaviour which is counter-intuitive and in clear contrast with the Einstein relation for systems in thermal equilibrium \cite{einstein1905}. $\langle \mathbf{v} \rangle$ is gradually diminishing with temperature increase. Parameters are the same as in Fig. \ref{fig2}.}
	\label{fig3}
\end{figure}
\section{Deterministic system}
The studied model possesses a five-dimensional parameter space $\{m, a, \omega, \varphi, Q\}$ which is too complex to analyse numerically in a systematic manner even with the help of our innovative computational methods. However, recently \cite{spiechowicz2015pre} we reported the remarkable regime where the diffusion coefficient $D$ exhibits non-monotonic dependence on temperature $\theta$ leading to an intriguing phenomenon of thermal noise suppressed diffusion. It corresponds to the following set of parameters $\{m = 6$, $a = 1.899$, $\omega = 0.403$, $\varphi = \pi/2\}$. Because noise assisted dynamics described by Eq. (\ref{model}) can be a repercussion of its deterministic properties as the first step we analyse the noiseless case $Q = 0$ (for the detailed discussion see Ref. 36). For this particular regime the system is non-chaotic and possesses three coexisting attractors. 
 The corresponding structure of basins of attraction for the asymptotic long time velocity $\mathbf{v}(t)$ is shown in Fig. \ref{fig2} (which is reproduced from our previous paper \cite{spiechowicz2016scirep}). The red and blue sets consists of all initial conditions $\{x(0), v(0)\}$ evolving to the running states with either positive $v_+ = \mathbf{v}(t) \approx 0.4$ and negative $v_- = \mathbf{v}(t) \approx -0.4$ velocity, respectively. The green colour marks the locked states $v_0 = \mathbf{v}(t) \approx 0$. 
There are three classes of trajectories corresponding to these three states: $x(t)\sim 0.4 t$, $x(t) \sim -0.4 t$ and $x(t) \sim 0$.  As a consequence superdiffusion occurs with  $\langle \Delta x^2(t) \rangle \sim t^2$. 
 Adding thermal noise causes a stochastic  dynamics which destabilizes the attractors and leads to random transitions between its coexisting basins of attraction. They play an analogous role to potential wells in equilibrium systems. Transitions between the running and/or locked states may generate the diffusion process and transient anomalous diffusion can be observed for low to moderate temperature regimes \cite{spiechowicz2016scirep}.
\begin{figure}[t]
	\centering
	\includegraphics[width=1.0\linewidth]{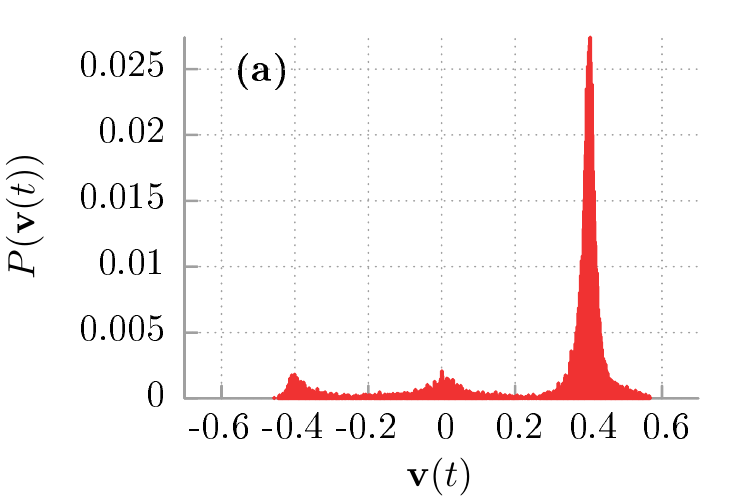}\\
	\includegraphics[width=1.0\linewidth]{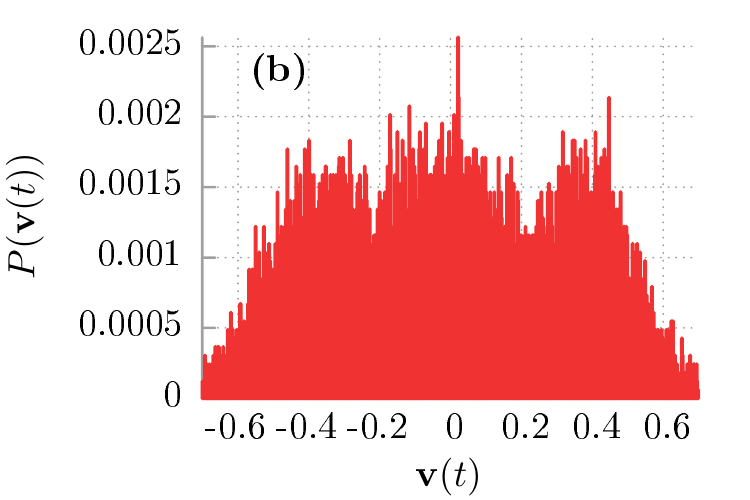}
	\caption{The probability distributions $P(\mathbf{v}(t))$ of the individual asymptotic long time period averaged velocity $\mathbf{v}(t)$ ($t = 10^4 \times T)$ are presented for two different values of thermal noise intensity corresponding to the maximal and (panel (a), $Q = 0.0045$) and minimal diffusion coefficient (panel (b), $Q = 0.76$). At small noise strength, as in panel (a), the fine structure of distribution is visible. High noise intensity presented in panel (b) leads to disappearance of this structure and flattening of the velocity distribution. Other parameters are the same as in Fig. \ref{fig2}.}
	\label{fig4}
\end{figure}
\section{Diffusion vs temperature}
In Fig. \ref{fig3} we present the dependence of the diffusion coefficient $D$ on the thermal noise intensity $Q \propto \theta$ which is proportional to temperature $\theta$ of the system. The diffusion coefficient $D$ behaves there in a nonlinear and non-monotonic manner. For low temperature $D$ initially increases, passes through its local maximum for $Q \approx 0.0045$ and next starts to decrease reaching its  minimum for $Q \approx 0.76$. For larger temperature,  $D$ monotonically increases and becomes strictly proportional to $Q$ (not depicted). The revealed behaviour stays in clear contrast to the Einstein relation for systems at thermodynamic equilibrium \cite{einstein1905} as well as to other known formulas, e.g. Arrhenius-type dependence for the diffusion of a Brownian particle in periodic potentials \cite{htb1990,goychuk2014}. In Fig. \ref{fig3} we present also the directed transport velocity $\langle \mathbf{v} \rangle$. This quantity is gradually diminishing with temperature increase and approaches zero  for sufficiently large $Q > 0.76$. It is so because then the deterministic forces in the right hand side of Eq. (\ref{model}) become progressively negligible in comparison to the thermal random force  $\xi(t)$ and the model describes a free Brownian particle.
\begin{figure}[t]
	\centering
	\includegraphics[width=1.0\linewidth]{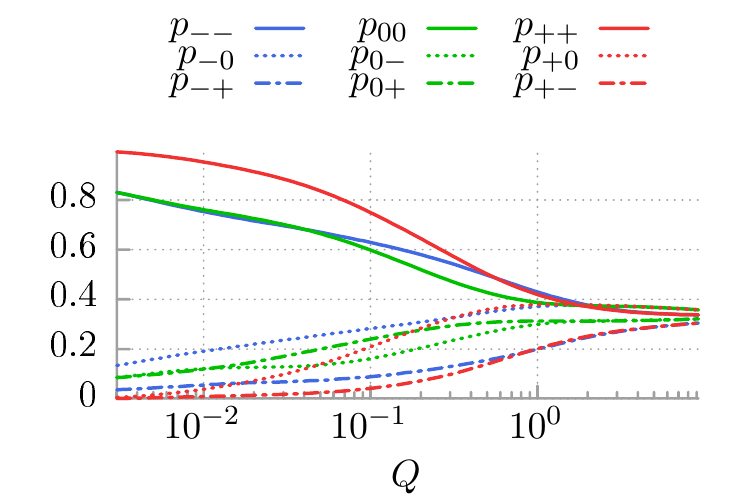}
	\caption{All transition probabilities between the three observed states: the minus $v_- = -0.4$, the zero $v_0 = 0$ and the plus $v_+ = 0.4$ solution. In the regime $Q = 0.0045$ corresponding to the maximal diffusion coefficient there are significant differences between them which disappear as temperature is increased. Other parameters are the same as in Fig. \ref{fig2}.}
	\label{fig5}
\end{figure}
\section{Velocity distribution}
In order to unravel the mechanism standing behind this discussed peculiar diffusive behaviour let us first have a look at the probability distributions $P(\mathbf{v}(t))$ of the individual asymptotic long time period averaged velocities $\mathbf{v}(t)$. These are depicted in \mbox{Fig. \ref{fig4}}. Panels (a) and (b) correspond to the maximal and minimal diffusion coefficient for $Q = 0.0045$ and $Q = 0.76$, respectively. In panel (a) a fine asymmetric structure of the probability distribution is visible. Most of the particles travel with a positive velocity corresponding to $v_+ = 0.4$. Although they are much less pronounced one can detect in this panel two satellite peaks describing the locked state $v_0 = 0$ and the negative running solution corresponding to $v_- = -0.4$. The three state structure in the presented regime is inherited from the deterministic counterpart of the dynamics when three non-chaotic attractors coexists: the running state with either positive or negative velocity $v_\pm = \pm 0.4$ and the locked state $v_0 = 0$, see also Fig. \ref{fig2}. An increase of thermal noise intensity leads to disappearance of this fine structure and flattening of the velocity distribution. This common feature is visualized in panel (b). Moreover, then the distribution becomes nearly symmetric with the center at zero having gained weight. The last fact agrees well with the dependence of the directed transport velocity $\langle \mathbf{v} \rangle$ on temperature presented in Fig. \ref{fig3}. Therefore we can see that the observed peculiar diffusive behaviour must be related to the exposed specific structure of the velocity probability distribution, in particular to the transitions between the observed three states.
\begin{figure}[t]
	\centering
	\includegraphics[width=0.9\linewidth]{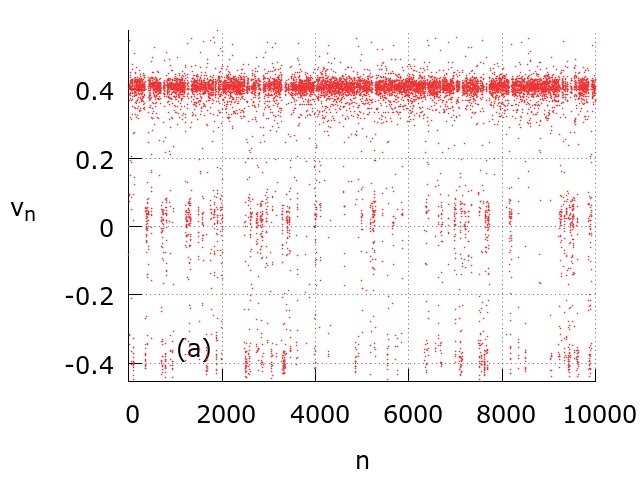}\\
	\includegraphics[width=0.9\linewidth]{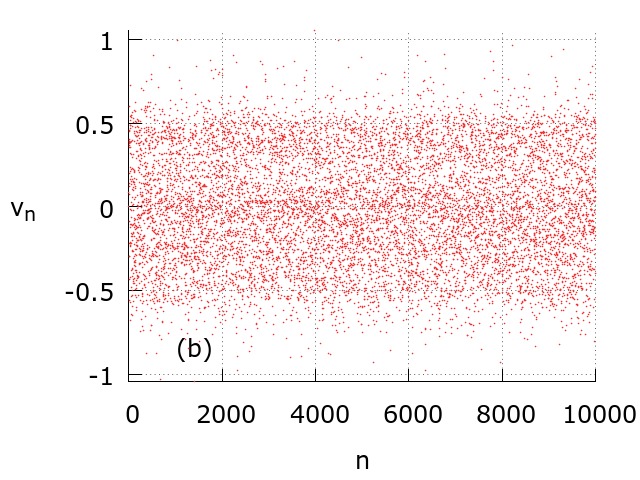}
	\caption{Representative single trajectory of the period averaged velocity $\mathbf{v}_n$ is presented for two different temperatures $Q = 0.0045$ (panel (a)) and $Q = 0.76$ (panel (b)) corresponding to the maximal and minimal diffusion coefficient, respectively. Each red dot represents the period averaged velocity $\mathbf{v}_n$ for the given period $n$. In the case of maximal diffusion coefficient (panel (a)) the particle stays predominantly in the plus state $v_+$ and from time to time takes a long lasting excursion to the other states. When temperature is increased (panel (b)) and the diffusion is decreased the particle jumps randomly between the three state without any apparent regularity.}
	\label{fig6}
\end{figure}
\section{Transition probabilities} 
The three maxima in the distribution of the period averaged velocity concentrate around the deterministically coexisting attractors. This fact allows us to construct a three-state stochastic process with jumps between states induced by thermal equilibrium fluctuations. Our goal is not to formulate and analyse this approximation in detail but rather to understand its basic properties. To this aim, we consider the influence of temperature variation on all transition probabilities between three states: $v_+ = 0.4$,  $v_0 = 0$ and $v_- = -0.4$. The notation $p_{++}$ is introduced for the conditional probability to remain staying in the plus state $v_+ \to v_+$, next $p_{+-}$ denotes the conditional probability of a transition between opposite running states $v_+ \to v_-$ and $p_{+0}$ is the conditional probability of a transition between the plus and the zero state, $v_+ \to v_0$. This naming convention is analogous for the other six conditional probabilities $\{p_{00}, p_{0+}, p_{0-}, p_{--}, p_{-0}, p_{-+}\}$. These characteristics are depicted in Fig. \ref{fig5} as a function of temperature of the system. We note that in the regime $Q = 0.0045$ corresponding to the maximal diffusion coefficient differences between the probabilities are significant. In particular, all probabilities for the particle to survive in each of the states $p_{++}$, $p_{00}$ and $p_{--}$ are large. The particle most likely resides in the running solution corresponding to the plus velocity $v_+ = 0.4$ which agrees well with the asymptotic probability distribution $P(\mathbf{v}(t))$ presented in \mbox{Fig. \ref{fig4} (a)}. However, now we can see that once the particle leaves this state for the zero or minus solution it will remain there also for a long time thus increasing spread of the trajectories and in turn also the diffusion coefficient. By analysing this panel we are able to identify the most probable transition cycles for the period averaged velocity $\mathbf{v}(t)$ in the low to moderate temperature regime. They are $v_+ \to v_0 \to v_+$ or $v_+ \to  v_- \to v_0 \to v_+$. As thermal noise intensity $Q$ is increased the differences between the transition probabilities are gradually disappearing and they are become equivalent. This fact again agrees with the form of the asymptotic probability distribution $P(\mathbf{v}(t))$ shown in Fig. \ref{fig4} (b). Then the particle frequently changes its state, however, the residence times for each of them are not sufficiently large to significantly modify spread of the trajectories and therefore the diffusion coefficient is smaller.
\begin{figure}[t]
	\centering
	\includegraphics[width=1\linewidth]{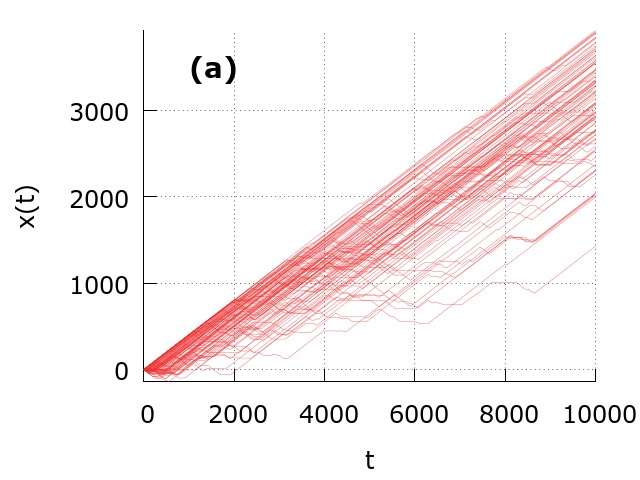}
	\includegraphics[width=1\linewidth]{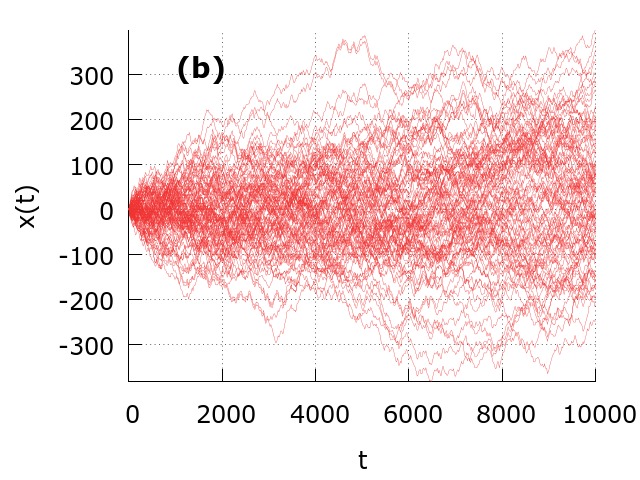}
	\caption{Typical sample trajectories of the Brownian particle coordinate $x(t)$ are depicted for different temperature of the system $Q = 0.0045$ (panel (a)) and $Q = 0.76$ (panel (b)). In panel (a) transition between the states $v_- = -0.4$, $v_0 = 0$ and $v_+ = +0.4$ are clearly visible. The long lasting periods of motion corresponding to the zero $v_0$ and the minus $v_-$ solutions lead to enlargement of the diffusion coefficient.}
	\label{fig7}
\end{figure}
\section{Spread of trajectories}
Now, in order to better visualize this qualitative mechanism we take a closer look at trajectories of the Brownian motor. We first study the time evolution of the period averaged velocity $\mathbf{v}(t)$. In \mbox{Fig. \ref{fig6}} we present an illustrative single trajectory of this observable. Each dot depicts the period averaged velocity $\mathbf{v}(t)$ corresponding to the given period $n$ of the driving, i.e. $\mathbf{v}_n = (1/T) \int_{nT}^{(n+1)T} ds\, \dot{x}(s)$. Panel (a) pertains to the maximal diffusion coefficient for temperature $Q = 0.0045$. One can evidently notice that during the time evolution the particle stays predominantly in the plus state $v_+$. However, from time to time it takes an excursion lasting for several dozens of the ac-driving periods to either the locked state $v_0$ or the running state $v_-$ with the  opposite direction. This last situation particularly enlarges the spread of  trajectories. On the other hand, in the case of high temperature $Q = 0.76$ and simultaneously minimal diffusion coefficient (panel (b)) the particle constantly jumps between these three states. While the difference between the period averaged velocities $v_n$ in two consecutive periods may be even greater than in the previous case it simply does not last sufficiently long to alter the overall spread of trajectories significantly.

Last but not least, in Fig. \ref{fig7} we depict various typical sample trajectories of the Brownian particle coordinate $x(t)$. Panel (a) relates to the regime of maximal diffusion coefficient for which temperature $Q = 0.0045$. Clearly, one can denote there transitions between the three states $v_- = -0.4$, $v_0 = 0$ and $v_+ = +0.4$. Moreover, the long lasting periods of motion corresponding to the zero $v_0$ and the minus $v_-$ solutions significantly increase the overall spread of trajectories. Most of the trajectories compactly follow the plus $v_+$ state but from time to time the particle jumps onto the other solution and stays there for many subsequent periods. This pictures agrees well with our previous statements deduced on the basis of Fig. \ref{fig5} and \ref{fig6} (a). The situation is very different in panel (b) corresponding to the minimal diffusion coefficient for which temperature $Q = 0.76$. There typical Wiener process-like trajectories of infinitely many turns back and forth are observed. Moreover, most of the trajectories oscillate around zero and there is no finite drift. Their overall spread is also much smaller.   Note that scales of the vertical axes in panels (a) and (b) are different and spread of trajectories for time $t=10 000$ is more than 2000 in panel (a) and about 700 in panel (b).  
\section{Summary}
In this work we investigated the diffusion in the archetypal model of the Brownian motor, i.e the Brownian particle moving in an asymmetric, one-dimensional periodic landscape of a ratchet type and subjected to an external unbiased time periodic force. We explain the mechanism standing behind the recently communicated anomaly \cite{spiechowicz2015pre} whereby the diffusion coefficient is non-monotonic function of temperature of the system. If temperature is increased the diffusion constant describing the spreading rate of the particle cloud first increases until it reaches a local maximum and then decreases until a minimum is hit. Further increase of temperature brings  a continuous growth of the diffusion coefficient which eventually becomes proportional to it.

As the mechanism for this counter-intuitive diffusive behaviour  we propose the temperature dependence of transitions between certain regions in the phase space dynamics of the motor. The latter contains in the deterministic case three non-chaotic attractors. Two of them are running solutions corresponding to the opposite velocity $v_\pm = \mathbf{v}(t) = \pm 0.4$ and another one is the locked state $v_0 = \mathbf{v}(t) = 0$. Transition probabilities between these three states vary with temperature leading to the mentioned dependence of the diffusion coefficient on temperature. In particular, when the diffusion is maximal the particle most likely resides in the $v_+$ state. However, at the same time the transition probabilities for the particle to stay in the remaining states $v_-$ and $v_0$ are significantly larger than the rest of rates. Therefore once the particle jumps from the state $v_+$ onto $v_-$ or $v_0$ it will remain there for a long time thus increasing spread of the trajectories and in turn also the diffusion coefficient. As temperature is increased the discrepancies between the transitions probabilities are gradually disappearing and they are become equivalent. Then the particle jumps between the three states very frequently and the diffusion coefficient is smaller.

A similar behaviour has been recently detected in symmetric periodic systems (SPS) 
\cite{spiechowicz2016njp}. However,  both systems are radically different.  In ratchet systems like considered here, one can observe directed transport which is quantified by the asymptotic  avegarged  velocity $\langle \mathbf{v} \rangle \neq 0$ while for SPS  there is no directed transport, i.e. the long-time averaged   velocity is zero,  $\langle \mathbf{v} \rangle = 0$. 
In the deterministic limit of zero temperature $\theta=0$,  the system analyzed here is  in a non-chaotic regime while the SPS has been studied in a chaotic state indicating deterministic diffusion \cite{spiechowicz2016njp}.  The non-monotonic temperature dependence of diffusion was inseparably related to a few {\it unstable} periodic orbits embedded into a chaotic attractor together with thermal noise induced dynamical changes upon varying temperature. It should be contrasted with the operational regime of our present setup which is clearly non-chaotic  
with three coexisting attractors and  three disjoint basins of attractions  which are mutually inaccessible. In consequence, the deterministic system is non-ergodic in the present case and transient anomalous diffusion can take place in the noisy system. Therefore this paper presents an alternative mechanism standing behind this fascinating phenomenon. We note that it may occur also in symmetric systems possessing multiple deterministically coexisting attractors provided that the symmetry condition is still satisfied $\langle \mathbf{v} \rangle = 0$. Moreover, the non-monotonic dependence of $D$ on temperature has lately been studied in  the system of  Brownian inertial particles moving in biased washboard potentials \cite{lindner2016}.   
In such systems, locked and running states (but only in one direction)   exist. The authors proposed a two-state theory incorporating the transition rates between the locked and running states  which reproduces this non-monotonic behavior.

Finally, we remind that the Langevin equation (\ref{model}) has its physical realization as  an asymmetric superconducting quantum
interference device (SQUID) subjected to a time-periodic current and pierced by an external magnetic flux. This asymmetric SQUID is formed by a superconducting loop with two resistively and capacitively shunted Josephson junctions in the left arm
and only one in the right arm \cite{spiechowicz2014prb,spiechowicz2015njp}. The quasi-classical dynamics of such systems  is well known in the literature as the Stewart-McCumber model \cite{blackburn2016}. Therefore our results can readily be experimentally tested with an accessible setup.  We want to mention that in the previous paper \cite{spiechowicz2015chaos}, we have studied diffusion properties of the Josephson phase in relation to quality of directed transport and in the chosen region of system parameters the diffusion coefficient, contrary to what is revealed this paper,  is monotonically increasing function of temperature.

In view of the widespread applications of Brownian motor setups and ratchet devices our research may bring along impact for further development of a working principle of a nanomotor operating on smallest scales occurring in diverse areas of science. Among others, one of the most promising aims is to harvest such nanomotors to convey powerful and efficient mechanisms that reach the ultimate effectiveness of biological systems which are responsible for the emergence and existence of the phenomenon of Life.

\section*{Acknowledgments}
The work  was supported in part by the MNiSW program via a Diamond Grant (J.S.) and the Grant NCN 2015/19/B/ST2/02856 (J. {\L}.). 


\begin{thebibliography}{99}
\bibitem{rogers2003} E. M. Rogers, \textit{Diffusion of Innovations}, (Free Press Simon and Schuster, New York, 2003)
\bibitem{einstein1905} A. Einstein, \textit{Ann. Phys.} \textbf{17}, 4549 (1905)
\bibitem{smoluchowski1906} M. Smoluchowski \textit{Ann. Phys.} \textbf{21}, 756 (1906)
\bibitem{chowdhury2013} D. Chowdhury, \textit{Phys. Rep.} \textbf{529}, 1 (2013)
\bibitem{spiechowicz2013jstatmech} J. Spiechowicz, J. \L uczka and P. H\"anngi, \textit{J. Stat. Mech.}, P02044 (2013)
\bibitem{spiechowicz2014pre} J. Spiechowicz, P. H\"anggi and J. \L uczka, \textit{Phys. Rev. E} \textbf{90}, 032104 (2014)
\bibitem{spiechowicz2016jstatmech} J. Spiechowicz, J. \L uczka \& L. Machura, \textit{J. Stat. Mech.} 054038 (2016)
\bibitem{bressloff2013} B. C. Bressloff \& J. M. Newby, \textit{Rev. Mod. Phys.} \textbf{85}, 135 (2013)
\bibitem{hanggi2009} P. H\"anggi and F. Marchesoni, \textit{Rev. Mod. Phys.} \textbf{81}, 387 (2009)
\bibitem{hoffmann2016} P. Hoffmann, \textit{Rep. Prog. Phys.} \textbf{79}, 032601 (2016)
\bibitem{metzler2014} R. Metzler, J. H. Jeon, A. G. Cherstvy and E. Barkai, \textit{Phys. Chem. Chem. Phys.} \textbf{16}, 24128 (2014)
\bibitem{zaburdaev2015} V. Zaburdaev, S. Denisov and J. Klafter, \textit{Rev. Mod. Phys.} \textbf{87}, 483 (2015)
\bibitem{lindner2001} B. Lindner, M. Kostur and L. Schimansky-Geier, \textit{Fluct. Noise Lett.} \textbf{1}, R25 (2001)
\bibitem{heinsalu2004} E. Heinsalu, R. Tammelo  and O. Teet, \textit{Phys. Rev. E} \textbf{69}, 021111 (2004)
\bibitem{lindner2016} B. Lindner \& I. M. Sokolov, \textit{Phys. Rev. E} \textbf{93}, 042106 (2016)
\bibitem{dan2002} D. Dan \& A. M. Jayannavar, \textit{Phys. Rev. E} \textbf{66}, 041106 (2002)
\bibitem{garcia2014} R. Salgado-Garcia, \textit{Phys. Rev. E} \textbf{90}, 032105 (2014)
\bibitem{vlassiouk2007} I. Vlassiouk  and Z. S. Siwy, \textit{Nano Lett.} \textbf{7}, 552 (2007)
\bibitem{serreli2007} V. Serreli, C. F. Lee, E. R. Kay and D. A. Leigh, \textit{Nature} \textbf{445}, 523 (2007)
\bibitem{mahmud2009} G. Mahmud \textit{et al.}, \textit{Nat. Phys.} \textbf{5}, 606 (2009)
\bibitem{costache2010} M. V. Costache \& S. O. Valenzuela,
\textit{Science} \textbf{330}, 1645 (2010)
\bibitem{drexler2013} C. Drexler \textit{et al.}, \textit{Nat. Nanotechnol.} \textbf{8}, 104 (2013)
\bibitem{spiechowicz2014prb} J. Spiechowicz, P. H\"anggi \& J. \L uczka, \textit{Phys. Rev. B} \textbf{90}, 054520 (2014)
\bibitem{spiechowicz2015njp} J. Spiechowicz and J. \L uczka, \textit{New J. Phys.} \textbf{17}, 023054 (2015)
\bibitem{spiechowicz2015chaos} J. Spiechowicz and J. \L uczka, \textit{Chaos} \textbf{25}, 053110 (2015)
\bibitem{roche2015} B. Roche \textit{et al.}, \textit{Nat. Commun.} \textbf{6}, 6738 (2015)
\bibitem{grossert2016} C. Grossert \textit{et al.}, \textit{Nat. Commun.} \textbf{7}, 10440 (2016)
\bibitem{guo2014} M. Guo , H. Gelman and M. Gruebele, \textit{PLoS ONE} \textbf{9}, e113040 (2014)
\bibitem{schuring2002} A. Sch{\"u}ring, S. M. Auerbach, S. Fritzsche and R. Haberlandt, \textit{J. Chem. Phys.} \textbf{116}, 10890 (2002)
\bibitem{tung2016} W. S. Tung \textit{et al.}, \textit{ACS Macro Lett.} \textbf{5}, 735-739 (2016)
\bibitem{ganshin1999} A. N. Gan'shin \textit{et al.},
\textit{Low Temp. Phys.} \textbf{25}, 259 (1999)
\bibitem{eltsov2006} V. B. Eltsov \textit{et al.}, \textit{Phys. Rev. Lett.} \textbf{96}, 215302 (2006)
\bibitem{iubini2015} S. Iubini, O. Boada, Y. Omar and F. Piazza, \textit{New J. Phys.} \textbf{17}, 113030 (2015) 
\bibitem{lee2015} Ch. K. Lee, J. Moix and J. Cao, \textit{J. Chem. Phys.} \textbf{142}, 164103 (2015)
\bibitem{spiechowicz2015pre} J. Spiechowicz and J. \L uczka, \textit{Phys. Rev. E} \textbf{91}, 062104 (2015); See also article addendum at arXiv:1506.00105
\bibitem{spiechowicz2016scirep} J. Spiechowicz, J. \L uczka and P. H\"anggi, \textit{Sci. Rep.} \textbf{6}, 30948 (2016)
\bibitem{jung1993} P. Jung, \textit{Phys. Rep.}, 1993, \textbf{234}, 175.
\bibitem{spiechowicz2015cpc} J. Spiechowicz, M. Kostur and {\L}. Machura, \textit{Comp. Phys. Commun.} \textbf{191}, 140 (2015)
\bibitem{htb1990} P. H\"anggi, P. Talkner and M. Borkovec \textit{Rev. Mod. Phys.} \textbf{62}, 251 (1990); see Sect. VII therein
\bibitem{goychuk2014} I. Goychuk \& V. O. Kharchenko, \textit{Phys. Rev. Lett.} \textbf{113}, 100601 (2014)
\bibitem{spiechowicz2016njp} J. Spiechowicz, P. Talkner, P. H\"anggi and J. \L uczka, \textit{New J. Phys.} \textbf{18}, 123029 (2016)
\bibitem{blackburn2016} J. A. Blackburn, M. Cirillo and N. Gronbech-Jensen, \textit{Phys. Rep.} \textbf{611}, 1 (2016)
\end{thebibliography}
\end{document}